\documentclass[epj]{svjour}
\usepackage{graphics}
\usepackage[dvips]{color}
\begin{document}

\title{Fine structure in the azimuthal transverse momentum correlations at
  $\sqrt{s_{NN}}=200$ GeV using the event shape analysis}

\author{Alejandro Ayala \and   Eleazar  Cuautle
  \and   Isabel Dom\'{\i}nguez
  \and  Antonio Ortiz \and  Guy Pai\'c}
\institute{Instituto    de    Ciencias
Nucleares,  Universidad  Nacional  Aut\'onoma  de  M\'exico,\\
  Apartado Postal   70-543,   M\'exico    D.F.   04510,   M\'exico.}
\date{}

\abstract{The experimental results on transverse momentum azimuthal hadron
  correlations at RHIC have opened a rich field for parton energy loss
  analysis in heavy-ion collisions. Recently, a considerable amount of work
  has been devoted to study the shapes of the ``away-side'' jet which exhibit
  an interesting and unexpected ``double hump'' structure not observed in the
  analogous treatment of $pp$ data. Driven by the possibility that the latter
  result might just mean that such structure exists already in the case of
  $pp$ collisions, but that its relative intensity could be small, here we use
  the Event Shape Analysis to show that it is possible to identify and select
  well defined event topologies in $pp$ collisions, among which, a double hump
  structure for the away-side jet emerges. Using two shape parameters, the
  sphericity in the transverse plane and the recoil to analyze a sample of PYTHIA
  generated $pp$ collisions at $\sqrt{s_{NN}}=200$ GeV, we show that this
  structure corresponds to two jets emitted in the backward
  hemisphere. Finally, we show that Q-PYTHIA qualitatively reproduces the
  decrease in the yield of dijet events and the increase of the double hump
  structure in the away side observed in heavy ion collisions. The
  implications for the treatment of parton energy loss in heavy-ion collisions
  are discussed. }\authorrunning{A. Ayala et. al} 
\titlerunning{Fine Structure in the azimuthal
  transverse momentum correlations at $\sqrt{s_{NN}}=200$ GeV...}  
\maketitle

\section{Introduction}\label{intro} 

The study of transverse momentum azimuthal correlations (further
referred to as azimuthal correlations) of high momentum particles has
considerably enhanced our understanding of the parton energy loss
mechanism in the dense matter created in the aftermath of a heavy ion
collision. The two-particle (dihadron) relative azimuthal angle,
$\Delta \phi$, correlation technique provides an alternative approach
for accessing the properties of jets.  Two classes of hadrons, trigger
and partner hadrons, typically from different $p_t$ ranges, are
correlated with each other. Jet properties are extracted on a
statistical basis from the $\Delta \phi$ distribution built of many
events. This approach overcomes problems due to background and limited
acceptance and allows the study of jets to be extended to low $p_t$
where soft processes dominate~\cite{Adler}.

However after the initial simple explanations to account for the jet
shapes the surprise came, namely, that for a given combination of
triggering momenta and associated particles, a double hump structure
was observed in the away-side associated particle
correlations~\cite{:2008cqb} and/or a broadening of the away-side
peak~\cite{wang}.

The observation brought about a large number of theoretical
explanations with widely different approaches. The existing
explanations are based on the assumption that we are indeed facing a
difference between the $pp$ and the heavy-ion cases. Among the
proposed explanations we can mention: a Mach-like structure (splitting
of the away-side peak)~\cite{Casadelrrey}, gluon Cerenkov-like
radiation models~\cite{Majumder}, the parton cascade model, the
Markovian parton-scattering model~\cite{Chiu}, and the color wake
model~\cite{chaudhuri}. One can ask, however, whether the above
mentioned characteristics of the away-side peak are already present in
$pp$ collisions, only being obscured by the smallness of their
intensity.

In the present work we investigate the shape of the away-side
correlation peak in collisions at $\sqrt{s_{NN}}=200$ GeV using an
approach based on the Event Shape Analysis (ESA). The questions we
want to answer are:

\begin{enumerate}
\item[1.-] Is it possible to identify and isolate events with
  topologies that resemble the double hump structure, using ESA in
  $pp$ collisions?
\item[2.-] What is the effect of the event structure on the width of
  the away peak?
\item[3.-] What is the effect of the Q-PYTHIA~\cite{qpythia} parton
  energy loss model on the event shape and away-side correlations?
\item[4.-] What are the theoretical implications of the results of the present
  analysis for parton energy loss consideration?
\end{enumerate}

The work is organized as follows: In Sec.~2 we present the basis for
the ESA. In Sec.~3 we show the results obtained for azimuthal
correlations for different parts of the event shape phase space

In Sec.~4 we analyze the effect of the recent Q-PYTHIA
generator on the azimuthal correlations. Finally in Sec.~5 we present
a theoretical discussion and an outlook of the results as well as the
conclusions.


\section{Event Shape Analysis}\label{eventshape}

The $e^+ - e^-$ experiments have introduced numerous event shape
variables to analyze their data. In hadronic collisions the same
approach is not possible due to the fact that the total event shape is
dominated by the beam axis. For that reason one defines event shape
variables in the plane transverse to the beam. For the sake of the
current work we limit ourselves to two parameters, Thrust $(T)$ and
Recoil $(R)$~\cite{Banfi:2004nk}. $T$ is defined as:

\begin{equation}
   T\equiv \underbrace{max}_{\overrightarrow{n}_T}
   \frac{\sum_{i}|\overrightarrow{q}_{\perp,i}\cdot\overrightarrow{n}_{T}|}
        {\sum_{i}|\overrightarrow{q}_{\perp,i}|},
\end{equation}
where the sum runs over all particles in the final state within the
acceptance, $\overrightarrow{q}_{\perp,i}$ represent the momentum
components transverse to the beam and $\overrightarrow{n}_{T}$ is the
transverse vector that maximizes the ratio. $T$ (the sphericity in the
transverse plane, $1-T$, closely related to $T$) will be equal to 1
(0) for a jet detected within the acceptance. On the other hand a
fully isotropic event will result in a sphericity of 0.5. $R$ is
defined as:

\begin{equation}
R\equiv \frac{|\sum_{i}\overrightarrow{q}_{\perp
    i}|}{\sum_{i}|\overrightarrow{q}_{\perp i}|},
\end{equation}
where again, the sum runs over all particles in the final state within
the acceptance. The necessity for this variable stems from the incomplete
acceptance we are faced with in the present application. As defined above, the
maximum value of the recoil will be for a jet event where only one jet
is detected within the acceptance.  In short the Recoil variable takes
care of momentum conservation.

\section{Results}\label{results}

\subsection{ The  event structure}\label{eventstructure}

\begin{figure}
\resizebox{0.5\textwidth}{!}{ \includegraphics{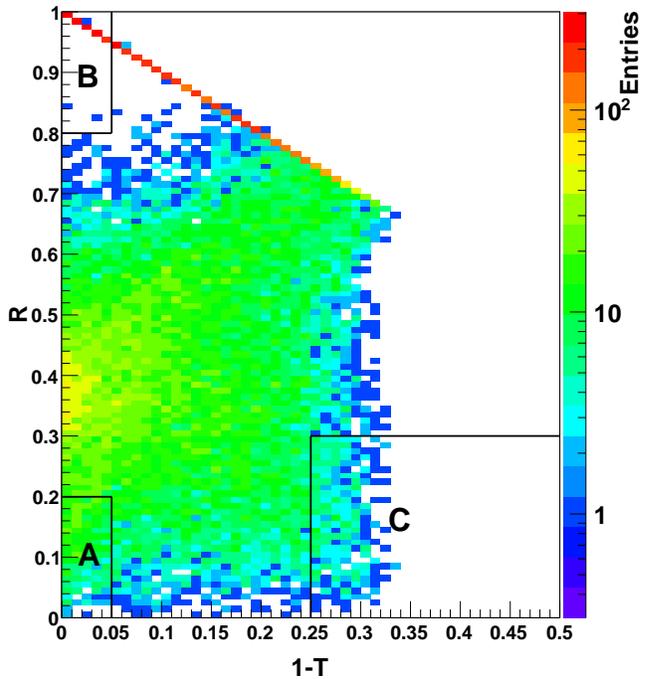}}
\caption{(Color online) $R$ vs $1-T$ for a minimum bias sample of $pp$
  collisions at $\sqrt{s_{NN}}=200$ GeV.}
\label{thrustmap200}
\end{figure}

Figure~\ref{thrustmap200}  shows the two-dimensional  distribution $R$
vs. $1-T$ obtained using PYTHIA  version 6.4.14 with minimum bias $pp$
collisions at  $\sqrt{s_{NN}}=200$ GeV. In the  simulation, we require
that all particles be  charged primaries and within the pseudorapidity
range of $|\eta|\leq1$. We also  require that the events have at least
3      particles     with      transverse      momentum     satisfying
$|\overrightarrow{q}_{\perp,i}|\geq0.8$ GeV/c.

The distribution shows the following main regions, labeled (A), (B) and (C):
\begin{itemize}
\item[(A)]The low $R$ and $1-T$ region which is known to correspond to
  dijets.
\item[(B)]The high $R$ and low $1-T$ region which is known to correspond to
  single jets (with the second jet being out of the acceptance).
\item[(C)]The high $1-T$ region which is known to correspond to the most
  isotropic events.
\end{itemize}

\begin{figure}
\resizebox{0.5\textwidth}{!}{
\includegraphics{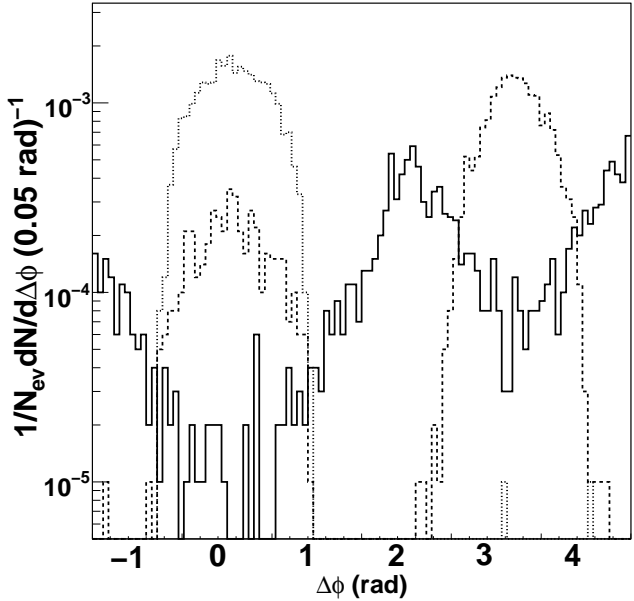}}
\caption{Azimuthal distribution for events in the different regions of
  the plot $R$ vs $1-T$. The highest momentum particle (leading
  particle) in the event is located at $\Delta\phi$ = 0 radians. Shown
  are:  events in region (A) (dashed line),
  events in region (B) (dotted line), events of region (C) (solid
  line).  Note that the peak corresponding to the leading particle is not shown.  }
\label{phi200gev}
\end{figure}

We now proceed to analyze the azimuthal distribution obtained from the
above  three  regions which  are  generated  by  choosing the  highest
momentum particle  in the event to be  at 0 radians. This  is shown in
Fig.~\ref{phi200gev},   where   the   near   side   peaks   are   well
distinguishable for the 3 plots while substantial differences occur in
the away-side. For  particles in region (A) we find  the usual form of
the away-side peak. For particles in region (B), as expected, there is
no sight of  a peak in the away-side. Finally  for particles in region
(C) we are confronted with a double hump structure. While the behavior
of the sample in region (C)  should in itself not be unexpected, since
it can be associated to a  gluon emission with a finite angle, like in
the  case of the  discovery of  the gluon~\cite{Barber:1979yr},  it is
interesting  that this  has not  been observed  in the  experiments at
RHIC, probably  due to the $p_t$  cut that was taken above 2 GeV/c.   One notes the
very distinct widths of the prominent  peaks in the case of taking the
whole inclusive data (not shown in the figure) instead of a subsection of back
to back jets only.

\begin{figure}
\resizebox{0.5\textwidth}{!}{ \includegraphics{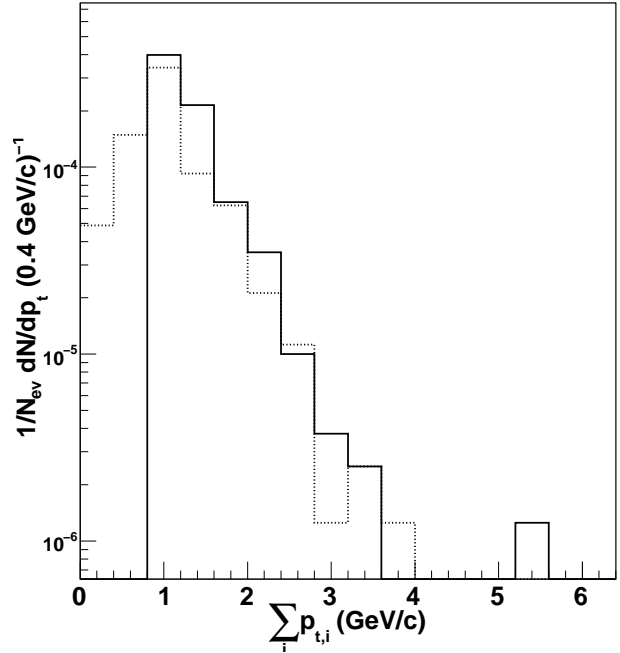}
} 
\caption{Transverse momentum spectra for the near side correlation
  peak (solid line) and the vector sum of the particle momenta in
  the away-side correlation peak (dotted line).  } \label{trijetpt}
\end{figure}

The main reason  for that lies in the fact  that the contribution from
region (C) can be described  as a mere broadening of the away peak
when all  events are taken and  is not readily  distinguishable in its
own right. Actually, one observes  a continuous broadening of the away
peak with increasing $1-T$ values. For the largest value of sphericity
the away peak transforms in  a double hump structure. We interpret the
structure as three jets emitted at 120 degrees from each other, as it
is illustrated in Fig. ~\ref{phi200gev}. In Fig.~\ref{trijetpt} we show
the spectrum for all particles  with $p_{t,i}>0.8$ GeV/c as a function
of the sum of the momenta  for all particles. We divided the azimuthal
correlations in  region (C)  in three parts:  one centered  in $\Delta
\phi$  around   zero,  the   other  two  around   the  two   humps  at
$\Delta\phi\sim  2.1$  radians and  $\Delta\phi\sim  4.2$  radians in  the
away-side correlations (to take into account the hump in the near-side
jet and the two humps of the away-side jet, respectively).

\begin{figure}
\resizebox{0.5\textwidth}{!}{ \includegraphics{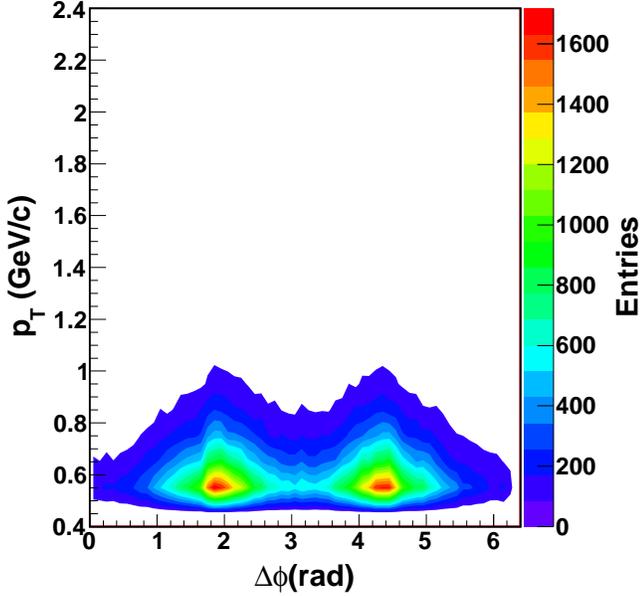}
} 
\caption{(Color online) Scatter plot of transverse momentum spectra versus 
$\Delta \phi$ for events of the region (C),  in the thrust map for minimum bias events}
 \label{ptvsphi}
\end{figure}

For each event,  a vector sum of the momenta for  the particles in the
away-side  peaks has  been performed.  The spectrum  thus  obtained is
compared  to the  corresponding  vector  sum of  the  momenta for  the
particles in the near-side peak. The agreement between both spectra is
reasonable,  pointing to  the fact  that the  three jet  assumption is
valid \cite{ortiz}.  A similar structure in the azimuthal correlations
for  particles in region  (C) of  Fig.~\ref{thrustmap200} can  be seen
even at lower beam energies.   The result is not surprising given that
the  $SPS$  experiments  have   already  reported  their  analysis  on
di-hadron   correlations  in  heavy-ion   collisions  ~\cite{ploskon}.
Figure~\ref{ptvsphi} shows  the scatter  plot of $p_t$  versus $\Delta
\phi$  distribution,  where one  observes  explicitly  that the  $p_t$
distributions exhibit  also a peak for the $p_t$  values that correspond
to   the    two   peaks    seen   in   azimuthal    correlation   (see
Fig.~\ref{phi200gev}). The absence of the  peak at zero radians is due
to our subtracting  of the leading particle and  therefore to the lack
of associated  particles around  it. In our  opinion, the  features of
Fig.~\ref{ptvsphi} speak also in  favor of the interpretation of these
two structures as corresponding to two jets in the away side.

\section {Azimuthal correlations using the Q-PYTHIA generator}
\label{azcorqpythia}

Confronted with the results obtained above we have tried to see what
are the predictions of a generator that has been recently released and that takes
 into account the
effects of parton energy loss in the medium simulating the effects to
be observed in heavy ion collisions, the Q-PYTHIA
generator~\cite{qpythia}. This is a Monte Carlo implementation in PYTHIA of
medium-induced glu\-on radiation. Medium effects are introduced through an
additive term in the vacuum splitting functions
\begin{equation}
P_{\rm tot} (z)=P_{\rm vac}(z)+\Delta P(z,t,\hat{q},L,E).
\end{equation}
where $\Delta P$ is the full splitting probability as given by the
medium induced gluon radiation spectrum~\cite{qpythia}. All the medium
information in $\Delta P$ is encoded in the product $n(\xi)\,
\sigma({\bf r}) \simeq \frac{1}{2}\,\hat{q}(\xi)\, {\bf r}^2$, where
$n$ is the time-dependent density of scattering centers and $\sigma$
is the strength of a single elastic scattering. With this
medium-modified splitting function, Q-PYTHIA computes the Sudakov form
factor and the evolution equation in the following manner: given a
parton coming from a branching point with coordinates
($t_{1}$,$x_{1}$), where $t_{1}$ is its virtuality and $x_{1}$ its
energy fraction, the algorithm calculates the coordinates
($t_{2}$,$x_{2}$) for the next branching. The shower begins with a
parton that faces the full length of the medium $L$, so the medium
effects on the probability of the first branching are evaluated at
$L$. The coherence length of the emitted gluon is then computed and
its next branching is evaluated at $L-l_{coh}$, where
$l_{coh}={2\omega/k_{t}^2}$ is the gluon formation time and $\omega$
and $k_{t}$ are the energy and transverse momentum (with respect to
the parent parton) of the emitted gluon, respectively. The process is
iterated to get the entire energy evolution.

\begin{figure}
\resizebox{0.5\textwidth}{!}{
  {\includegraphics{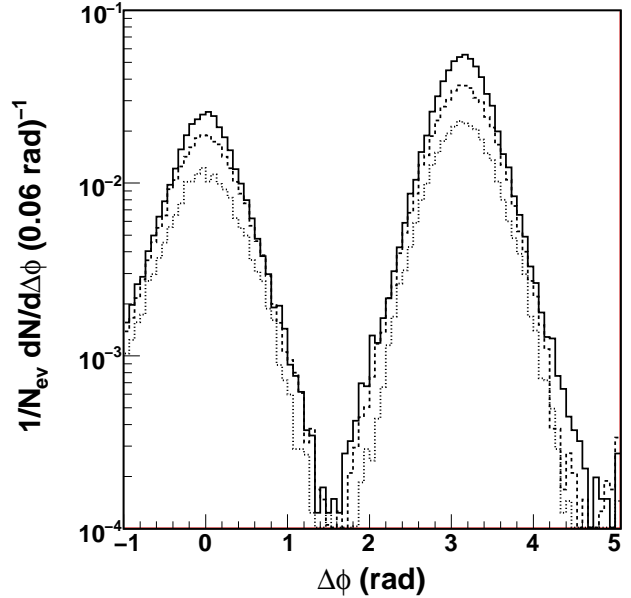}} }
\caption{Azimuthal correlations for particles in region (A) with
  p$_{t}^{hard} = 10$ GeV/c (solid line),
  Q-PYTHIA with    $\hat{q} = 1$ GeV$^2$/fm and $L =$ 6 fm (dashed line) 
  and $\hat{q} = 5$ GeV$^2$/fm and $L$ = 6 fm (dotted line),
  respectively. The chosen bounds for $1-T$ and $R$ are $1-T<0.05$ and
  $R<0.2$.} 
\label{deltaphi2jet}
\end{figure}

\begin{figure}
\resizebox{0.5\textwidth}{!}{
  {\includegraphics{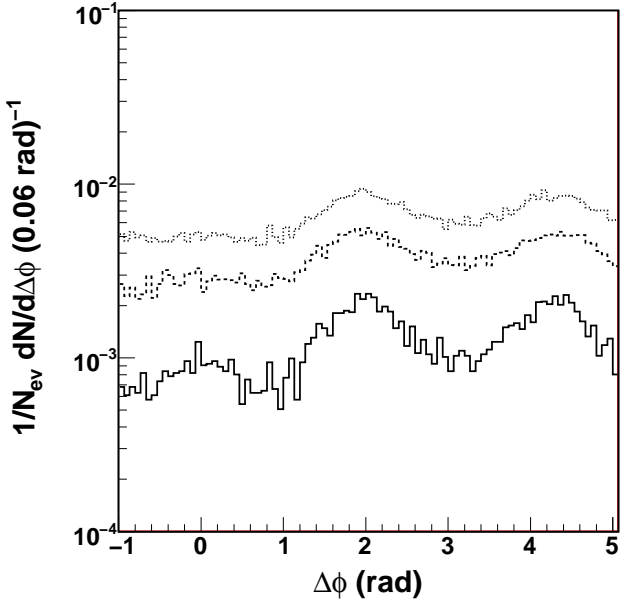}}} 
\caption{Azimuthal correlations for particles generated by PYTHIA in region (C) with
  p$_{t}^{hard} = 10$ GeV/c (solid line),  Q-PYTHIA
  with $\hat{q} = 1$ GeV$^2$/fm and $L =$ 6 fm (dashed line) 
  and $\hat{q} = 5$ GeV$^2$/fm and $L$ = 6 fm (dotted line),
  respectively. The chosen bounds for $1-T$ and $R$ are $1-T > 0.25$ and
  $R<0.3$.} 
\label{deltaphi3jet}
\end{figure}

In  Figs.~\ref{deltaphi2jet}  and~\ref{deltaphi3jet}  we  compare  the
effect  of  Q-PYTHIA on  the  event shape  using  a  sample of  events
generated with  a hard cutoff in $p_t$  of 10 GeV/c and  a lower $p_t$
cut on  the hadrons  of 0.5 GeV/c.  The curves obtained  correspond to
three distinct choices of parameters of Q-PYTHIA: A medium path length
of  6 fm  with  two choices  of  $\hat{q}$, 1  and  5 GeV$^2$/fm.  The
interest of the results lays in  the behavior of the dependence of the
yields in two different parts  of the $1-T$ vs $R$ distribution. While
for  low  values  of $R$  and  $1-T$  we  observe  a decrease  of  the
probability with  increasing $\hat q$, in  the case of  high values of
$1-T$  we observe  a  frank increase  of  the yield  in the  away-side
correlation with the double hump. These results remind us very much of
the  double hump  structures encountered  in the  RHIC  experiments in
collisions of heavy ions. In Fig.~7  we show the projection of the two
dimensional plot  on the $1-T$  axis for all  values of $R$  less than
0.8.  Again the  evolution with  rising values  of $\hat  q$ is  to be
noted.

\section{Discussion and Conclusions}\label{conclusions}

We have  shown that the ESA  allows to select  events with distinctive
features such as a dominant dijet structure and/or multijet structure,
by means of the use of two very simple variables: the sphericity $1-T$
and the recoil $R$. It is  important to note that the present analysis
allows  for the  isolation  of event  classes  even in  case they  are
rare. Very significant is the  observation of three jet events in $pp$
collisions. To  our knowledge the  existence of such  events, although
predicted by  theory, have not been  identified in the  RHIC data. The
implementation of Q-PYTHIA shows a very notable evolution of the event
shape  from dominant  dijet structure  to a  more  isotropic geometry,
ultimately enhancing significantly event structures with three prongs.

Recall that the  radiation of gluons is a  well established phenomenon
in  QCD  processes  involving  fast  partons. For  instance,  in  deep
inelastic scattering,  the observation of  events with three  jets was
one  of the  smoking  guns to  confirm  QCD as  the  theory of  strong
interactions~\cite{Barber:1979yr}.  The distinct  signature of a gluon
being emitted at  a finite angle with respect to  its parent parton is
the broadening of  the $p_t^2$ hadron distribution of  the jets beyond
their natural  spread in momentum  given by the  uncertainty principle
(Fermi motion). In this context, gluon radiation takes place in vacuum
and, whereas the emitted gluon spectrum peaks for small gluon energies
and emission angles, there is  a finite probability for gluon emission
at larger  angles. This probability  can even be computed  reliably in
perturbation  theory when the  virtuality of  the exchanged  photon is
large.

In a medium, induced gluon emission is usually understood as an effect
whereby  medium  partons  induce  small  momentum-transfer  collisions
producing  radiation  collinear  to  the direction  of  the  traveling
parton. This is a widely accepted scenario to describe the energy loss
of  a  fast parton  traveling  in  a  thermal parton  medium,  usually
referred to as the GLV model~\cite{Gyulassy}. When the energies of the
propagating parton and the medium partons are clearly distinguishable,
the  transferred  momentum from  the  parton  to  the medium  in  each
collision  is small  and small  angle gluon  emission is  the dominant
mechanism for induced radiative energy loss. It has even been reported
that within the GLV model, a broadening of the away-side jet should be
expected~\cite{Vitev}.

\begin{figure}
\resizebox{0.5\textwidth}{!}{%
\includegraphics{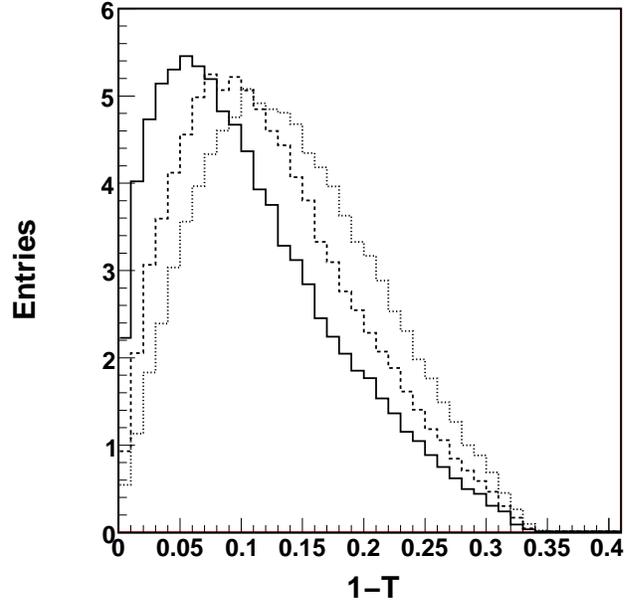}}
\caption{$1-T$ spectrum obtained for events with $R<$ 0.8 for
  particles with p$_{t}^{hard} = 10$ GeV/c (solid line),
  p$_{t}^{hard} = 10$ GeV/c and Q-PYTHIA with $\hat{q} = 1$ GeV$^2$/fm
  and $L =$ 6 fm (dashed line) and $\hat{q} = 5$ GeV$^2$/fm and $L$ = 6 fm
  (dotted line). }
\label{thrust06}
\end{figure}

The above  approximation for the description of  gluon emission should
be modified when  either the energy of the traveling  parton is not so
high or  the probability of  encountering other partons  with energies
not so small  as compared to the traveling  parton increases. In fact,
the PHENIX collaboration has reported  the appearance of a double hump
structure  in the away-side  jet as  the momentum  of the  leading and
corresponding  associated particles  diminishes~\cite{:2008cqb}. These
scenarios can  be achieved  in case  the parton has  an energy  not so
large  or by  the  increased probability  of  encountering partons  of
larger  mass/energy as compared  to the  thermal partons,  which could
happen early in  the reaction, that is, even  before thermalization is
achieved. As the medium becomes denser or dominated by mini-jets, this
picture should become important.

The fact that  a double hump structure for the  away-side jet has been
observed in heavy-ion  collisions and not in $pp$  collisions can thus
be  an indication  of a  medium  enhanced probability,  driven by  the
aforementioned scenarios, for a  process already present in the latter
reactions. Thus, the  need of a large intrinsic  $k_t\simeq 2.68$ GeV,
as  suggested in  Ref.~\cite{Adler2}  might just  reflect an  enhanced
probability for the  emission of gluons with large  angles relative to
the direction of the away-side  parton. We are currently pursuing such
possibilities to describe the effect~\cite{progress}.
 
One may  ask whether a comparison  of the ESA using  Q-PYTHIA and RHIC
data  can be  made, given  that  this event  generator introduces  jet
quenching   only  by   simulating  the   emission  of   radiation  and
disregarding  the background, whereas  in a  realistic situation  of a
heavy-ion  reaction  one knows  that  jets  come  along with  a  large
background.  In  this respect  it is important  to emphasize  that the
main goal  of this  work is  to point out  to the  interesting feature
shown by the  ESA, namely, the suppression of  di-jets and enhancement
of three jets in events with  large sphericity.  It is clear that in a
heavy-ion event, one should  introduce the appropriate cuts to recover
the  thrust  map  in  order  to  get  rid  of  the  underlying  event.
Therefore, in the  context of our work, the use  of Q-PYTHIA should be
taken only as an illustration of a possible effect of jet quenching in
a heavy-ion environment.

We have demonstrated the importance of event shape cuts to extract the
width of the away peak  in azimuthal correlations.  In the Monte Carlo
$pp$ analysis;  we have  observed a continuous  evolution of  the away
spectrum  starting from  a narrow  width of  the away-side  peak  to a
double  hump structure  for the  most  isotropic parts  at high  $1-T$
values.  We have  demonstrated the sensitivity of the  away peak width
to  the  part of  the  phase space  from  where  the correlations  are
extracted.

This study  suggests that  ESA may allow  a more detailed  analysis of
data. Finally, the predictions of an energy loss afterburner have been
shown. Surprisingly  this predicts an  enhancement of the  away double
hump correlation  structure with respect  to PYTHIA, opening  thus the
possibility  to  use the  amplitude  and shape  of  the  away peak  to
determine  $\hat q$.  The double  hump  effect observed  in heavy  ion
collisions  at RHIC  seems to  bear  some similarity  with the  shapes
observed here  in $pp$ collisions.   An analysis of data  in heavy-ion
collisions in a  similar way as proposed in this  work could shed some
additional light on the phenomenon.

\section*{Acknowledgments}
The authors express their gratitude to Dr. J.P. Revol for suggesting the
present investigation and for his judicious comments and to Dr. Leticia
Cunqueiro and Dr. Andreas Morsch for valuable discussions on the
implementation of Q-PYTHIA. Support for this work has been received in part by
DGAPA-UNAM under PAPIIT grants IN116008, IN115808 and IN116508 as well as by
the HELEN program.


\end{document}